\documentclass[sigconf, authorversion]{acmart}
\usepackage{float}
\usepackage{subcaption}
\usepackage{xspace}
\usepackage{url}
\usepackage[ruled,vlined,linesnumbered]{algorithm2e}
\usepackage{bm}
\usepackage{color}
\usepackage{soul}
\usepackage{balance}

\AtBeginDocument{%
  }

\copyrightyear{2024}
\acmYear{2024}
\setcopyright{acmlicensed}\acmConference[UbiComp Companion '24]{Companion of the 2024 ACM International Joint Conference on Pervasive and Ubiquitous Computing }{October 5--9, 2024}{Melbourne, VIC, Australia}
\acmBooktitle{Companion of the 2024 ACM International Joint Conference on Pervasive and Ubiquitous Computing (UbiComp Companion '24), October 5--9, 2024, Melbourne, VIC, Australia}
\acmDOI{10.1145/3675094.3678488}
\acmISBN{979-8-4007-1058-2/24/10}

\newcommand{\systemname}{\textit{AcousAF}\xspace}
\begin{document}

\title{AcousAF: Acoustic Sensing-Based Atrial Fibrillation Detection System for Mobile Phones}



\author{Xuanyu Liu}

\orcid{0009-0001-2825-311X}
\author{Haoxian Liu}
\orcid{0009-0004-7550-1325}
\affiliation{%
  \department{Research Institute of Trustworthy Autonomous Systems and Department of Computer Science and Engineering}
  \institution{Southern University of Science and Technology}
  \city{Shenzhen}
  \country{China}
}

\author{Jiao Li}
\orcid{0000-0002-3918-4922}
\author{Zongqi Yang}
\orcid{0009-0006-4191-8290}
\affiliation{%
  \department{Research Institute of Trustworthy Autonomous Systems and Department of Computer Science and Engineering}
  \institution{Southern University of Science and Technology}
  \city{Shenzhen}
  \country{China}
}

\author{Yi Huang}
\orcid{0009-0008-0039-4026}
\affiliation{
  \department{Department of Cardiology}
  \institution{Southern University of Science and Technology Hospital}
  \city{Shenzhen}
  \country{China}
}

\author{Jin Zhang}
\email{zhangj4@sustech.edu.cn}
\orcid{0000-0002-2674-0918}
\affiliation{%
  \department{Research Institute of Trustworthy Autonomous Systems and Department of Computer Science and Engineering}
  \institution{Southern University of Science and Technology}
  \city{Shenzhen}
  \country{China}
}
\authornote{Corresponding author}

\renewcommand{\shortauthors}{Xuanyu Liu et al.}

\ccsdesc[500]{Human-centered computing~Ubiquitous and mobile computing}

\keywords{Atrial Fibrillation Detection; Acoustic Sensing; Mobile Health}


\begin{abstract}
Atrial fibrillation (AF) is characterized by irregular electrical impulses originating in the atria, which can lead to severe complications and even death. Due to the intermittent nature of the AF, early and timely monitoring of AF is critical for patients to prevent further exacerbation of the condition. Although ambulatory ECG Holter monitors provide accurate monitoring, the high cost of these devices hinders their wider adoption. Current mobile-based AF detection systems offer a portable solution. However, these systems have various applicability issues, such as being easily affected by environmental factors and requiring significant user effort. To overcome the above limitations, we present \systemname, a novel AF detection system based on acoustic sensors of smartphones. Particularly, we explore the potential of pulse wave acquisition from the wrist using smartphone speakers and microphones. In addition, we propose a well-designed framework comprised of pulse wave probing, pulse wave extraction, and AF detection to ensure accurate and reliable AF detection. We collect data from 20 participants utilizing our custom data collection application on the smartphone. Extensive experimental results demonstrate the high performance of our system, with 92.8\% accuracy, 86.9\% precision, 87.4\% recall, and 87.1\% F1 Score.
\end{abstract}

\maketitle

\section{Introduction}
\label{s:intro}
Atrial fibrillation (AF), characterized by irregular electrical impulses originating in the atria, stands as the most prevalent arrhythmia \cite{AF-Intro}. Significantly, the presence of AF has a strong association with ischemic stroke \cite{AF-With-Stroke}, a severe condition capable of causing long-term disability or even death. Given the growing prevalence and the potential for severe complications, meticulous attention should be directed toward the early detection of AF.

However, diagnosing AF presents challenges in practice. Firstly, a considerable number of individuals with AF are asymptomatic, as evidenced by at least one-third of patients experiencing no symptoms \cite{silentAF}. This lack of symptoms discourages them from seeking cardiovascular examinations at hospitals. Additionally, AF tends to be intermittent, further complicating diagnosis. Furthermore, the clinical diagnosis of AF heavily relies on electrocardiogram (ECG), which demands costly medical facilities for data collection and specialized medical expertise for interpretation. 

To tackle the mentioned issues, many research efforts have focused on utilizing mobile devices, such as smartwatches and smartphones, for AF detection. A common approach in current studies involves using smartwatches equipped with photoplethysmography (PPG) sensors to gather pulse wave data for analysis \cite{EarlyDetectionAf, WristPPG-1, WristPPG-2, WristPPG-3, WristPPG-4, WristPPG-5, WristPPG-6}. While PPG-based methodologies offer cost-effective solutions, they are susceptible to variations in skin tone \cite{PPGSkinTone} and may cause discomfort to users due to emitted light, particularly when utilized during nighttime. Another method involves analyzing seismocardiography (SCG) or ballistocardiography (BCG) data captured by the inertial measurement unit (IMU) of smartphones \cite{SCGAF, SCG1, SCG2}. Given the prevalence of IMU sensors in smartphones, this approach is feasible. However, it necessitates users to recline and position the smartphone on their chests, demanding the user's effort and applicable only in specific scenarios. The third approach involves using mobile ECG devices to acquire heart activity data for analysis \cite{ECG2, ECG3, MobileECG-1, MobileECG-2, MobileECG-3, MobileECG-4, MobileECG-5, MobileECG-6}. Although ECG serves as the gold standard for AF detection, mobile ECG devices cannot provide data of the same quality as medical ECG devices, leading to diminished accuracy. What's more, the high cost of these devices prevents their wider adoption.

To overcome the above limitations, we propose \textit{AcousAF}, an AF detection system that leverages acoustic sensors (e.g., microphones or speakers) to capture the pulse waves from the wrists. \textit{AcousAF} offers a user-friendly solution for AF screening. When a user experiences discomfort, he or she can simply use their smartphone to perform a self-test anywhere, providing immediate results without the need for additional equipment. This approach addresses the issue of delayed diagnosis and treatment associated with AF's intermittent nature. It is particularly helpful in AF screening because many AF cases are intermittent and may not be detected during a hospital visit.

Although promising, two challenges underlie the design: 
\begin{itemize}
    \item The feasibility of using acoustic sensing to capture pulse waves via smartphones remains unexplored. While APG \cite{acous4} demonstrates the potential of using in-ear microphones to detect pulse waves, it leverages the unique structure of the ear canal, which differs significantly from the wrist. Therefore, we conducted a feasibility study and confirmed that it is possible to use acoustic sensing technologies to extract pulse waves from the wrist.
    \item Despite proving feasible, the highly sensitive nature of acoustic sensing presents challenges in extracting high-fidelity pulse waves. To address this, we propose a Channel Phase Response (CPR)-based pulse wave extraction approach. This method involves emitting a sine-wave probing signal and using CPR estimation to extract the pulse wave from the signal reflected off the wrist.
\end{itemize}

We summarize our contributions as follows:
\begin{itemize}
    \item To the best of our knowledge, \textit{AcousAF} is the first AF detection system implemented on COTS smartphones using acoustic sensing technologies. This innovative approach utilizes smartphones' built-in microphones and speakers for non-intrusive pulse wave monitoring, eliminating the need for additional sensors.
    \item We propose a well-designed pipeline that incorporates pulse wave probing, pulse wave extraction, and AF detection to facilitate accurate AF detection.
    \item We conduct experiments involving 20 subjects, comprising 6 individuals with AF and 14 healthy individuals. The results demonstrate the high performance of our system, achieving 92.8\% accuracy, 86.9\% precision, 87.4\% recall, and 87.1\% F1 Score.
\end{itemize}

\section{Feasibility Study}
\label{s: feasibility_study}

\begin{figure}[h]
  \centering
  \includegraphics[width=\linewidth]{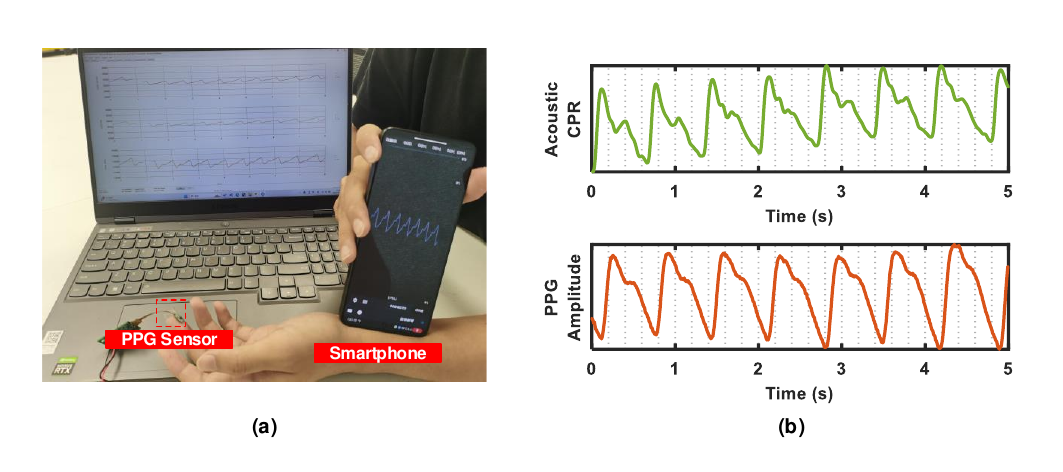}
  \caption{Feasibility study setup and result. \textbf{(a)}: Scenario of feasibility study. The PPG sensor is placed at the fingertip, and the smartphone is placed at the wrist above the radial artery. \textbf{(b)}: Comparison of PPG and Acoustic CPR. The CPR waveform shows a high similarity with the PPG waveform.}
  \label{f: feasibility}
\end{figure}

Reliable detection of atrial fibrillation requires the acquisition of high-fidelity pulse waves. The radial artery, located on the thumb side of the wrist, has strong pulsations, and these pulsations due to cardiac activity can be sensed through the fingertip by pressing on it with the finger. \textit{Similarly, if the speaker and microphone of a smartphone are pressed on the radial artery like a finger, can the tiny deformations of the skin surface caused by the heartbeat be sensed?} 

To answer this question, we conduct a feasibility study. In particular, we propose harnessing CPR estimation to analyze the tiny oscillations induced by the wrist's radial artery. A Redmi Note 13 Pro smartphone and a MAXM86161 PPG sensor are utilized for concurrent acquisition of acoustic and PPG data, as illustrated in Fig.~\ref{f: feasibility}(a). The study aims to compare the correlation between estimated CPR and PPG waveform. The smartphone, positioned above the wrist with its microphone facing the radial artery, emits an 18,000 Hz sine wave signal through the speaker and records the received signal for analysis. Subsequently, the estimated acoustic CPR is obtained using the algorithm detailed in Sec.~\ref{s: pulse_wave_extraction}. At the same time, the PPG sensor placed at the fingertip is recording PPG data as a comparison.

The experimental results, presented in Fig.~\ref{f: feasibility}(b), exhibit a strong correlation between the estimated CPR and PPG data, demonstrating the feasibility of extracting high-fidelity pulse waves from the wrist using smartphone speakers and microphones. This opens up the potential for detecting AF.

\section{System Overview}
\label{s: overview}
\begin{figure}[t]
  \centering
  \includegraphics[width=\linewidth]{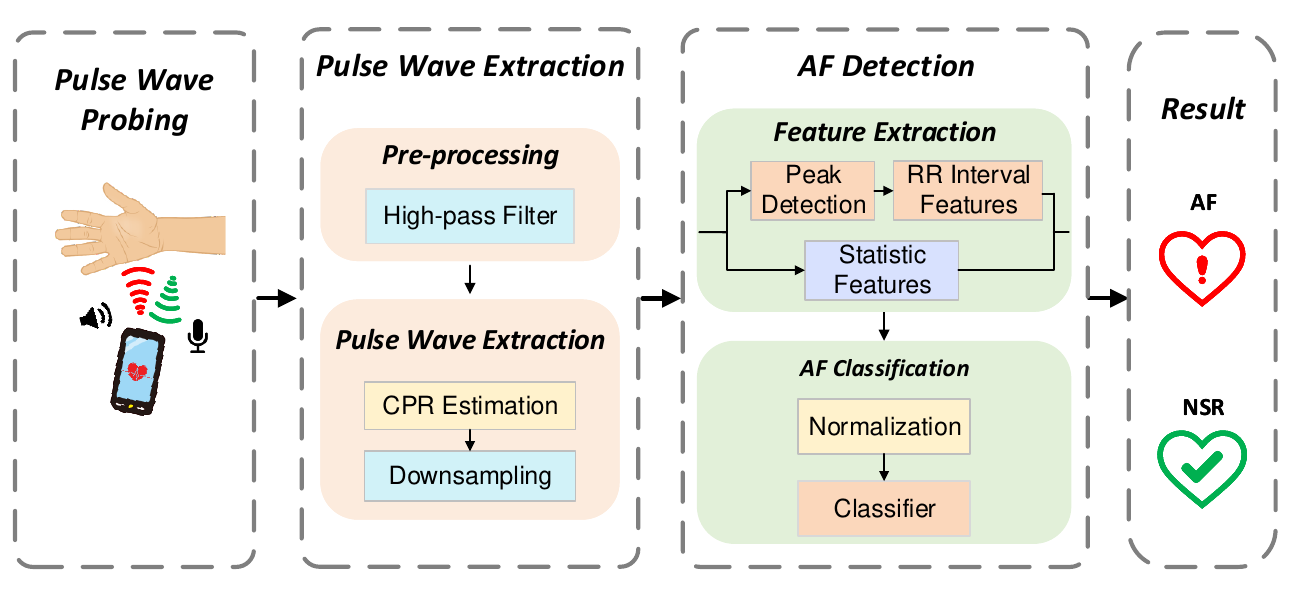}
  \caption{\systemname system overview.}
  \label{f: overview}
\end{figure}

Fig.~\ref{f: overview} illustrates the overview of our system, which is composed of three modules: \textbf{Pulse Wave Probing:} this module entails probing pulse waves from the wrist based on the speakers and microphones from the smartphones; \textbf{Pulse Wave Extraction:} noise of the acquired signals are removed and pulse waves are isolated utilizing our proposed CPR estimation method; \textbf{AF Detection:} feature extraction is performed to extract the RR interval features and statistic features from the extracted pulse waves, and AF classification is conducted for final AF detection.

\section{Pulse Wave Probing}
\label{s: pulse_wave_probing}

To accurately identify AF, the classical approach involves capturing pulse wave signals, which are intricately linked to cardiac activity. To facilitate this, our method involves monitoring minute wrist changes, utilizing a sine wave signal as a probing signal. As the transmitted signal traverses the wrist's skin and reaches the microphone, it is influenced by the underlying pulse wave information. \systemname algorithm exploits this by calculating the difference between the original and received signals, as detailed in Sec.~\ref{s: pulse_wave_extraction}. 

In aiming to implement accurate pulse wave probing, we also prioritize user experience. Our system sets the frequency of the probing signal at 18 kHz, which is imperceptible to most people. Additionally, the high-frequency probing signal reduces interference from daily noise, such as background sounds and conversations.
\section{Pulse Wave Extraction}
\label{s: pulse_wave_extraction}

\subsection{Pre-processing}
\label{ss: received_signal_pre-processing}
Upon reception by the microphone, the signal may be contaminated with ambient noise. To mitigate this, a bandpass filter is employed for each carrier, configured with a lower cutoff frequency of $f - 50$ Hz and a higher cutoff frequency of $f + 50$ Hz, where $f$ is the probing signal's frequency, ensuring a clean signal for further analysis.

\subsection{Pulse Wave Extraction}
\label{ss: pulse_wave_ext }
Pulse wave extraction lies at the core of cardiovascular monitoring for AF detection. This crucial process involves capturing and analyzing subtle variations in the pulse signal, which reflect the dynamic activity of the heart. To achieve this, we estimate CPR for an 18 kHz sine wave probing signal to extract the pulse wave. The acoustic signal transmitted from a smartphone speaker can be represented as $S(t) = \alpha \cos(2\pi f t)$, where $\alpha$ is the gain coefficient, and $f$ stands for the frequency of the probing signal, which is 18 kHz here. The signal transmitted through wrist skin and received by a smartphone speaker can be denoted as: 
\begin{align*}
S_{r}(t) &= A(t) \cos(2\pi f t - \theta_{c}(t) - \theta_{p}),
\end{align*}
where $A(t)$ is the amplitude of the received signal, $\theta_{c}(t)$ represents the phase response of the channel, i.e. CPR, and $\theta_{p}$ denotes the phase offset due to hardware delay and system noise. In most cases, $\theta_{p}$ can be considered as a constant and does not change with time.

Next, we utilize In-phase and Quadrature (I/Q) demodulation to extract the CPR for the probing signal. We first multiply the received signal $S_{r}(t)$ by $\cos(2\pi f t)$ to give a signal with the addition of one low-frequency component and multiple high-frequency components, which can be expressed as:
\begin{align*}
    S_{r}(t)\cos(2\pi f t) 
    &= A(t) \cos(2\pi f t - \theta_{c}(t) - \theta_{p})\cos(2\pi f t) \\
    &= \frac{1}{2}A(t)[\cos(\theta_{c}(t) + \theta_{p}) \\
    &\quad + \cos(4\pi f t - \theta_{c}(t) - \theta_{p})].
\end{align*}
In this equation, $\cos(4\pi f t - \theta_{c}(t) - \theta_{p})$ is naturally of high-frequency component, and $\cos(\theta_{c}(t) + \theta_{p})$ is the low-frequency component. After processing the signal with a low-pass filter, we can obtain the low-frequency component, which is called $I$ signal:
\[
I(t) = \frac{1}{2}A(t)\cos(\theta_{c}(t) + \theta_{p}).
\]
Similarly multiplying the received signal $S_{r}(t)$ by $\sin(2\pi f t)$ and passing it through a low-pass filter yields the $Q$ signal:\\
\[
Q(t) = \frac{1}{2}A(t)\sin(\theta_{c}(t) + \theta_{p}).
\]
Once $I$ and $Q$ signals are obtained, they are downsampled to 150 Hz to accelerate the computation for subsequent procedures. To obtain estimated CPR $\phi_{est}$, we only need to calculate the $arctan(\frac{Q}{I})$:
\begin{align*}
\phi_{est}(t) &= \arctan\left(\frac{Q}{I}\right) \\
           &= \arctan\left(\frac{\frac{1}{2}A(t)\sin(\theta_{c}(t) + \theta_{p})}{\frac{1}{2}A(t)\cos(\theta_{c}(t) + \theta_{p})}\right) \\
           &= \theta_{c}(t) + \theta_{p}.
\end{align*}

\begin{figure}
  \centering
  \includegraphics[width=\linewidth]{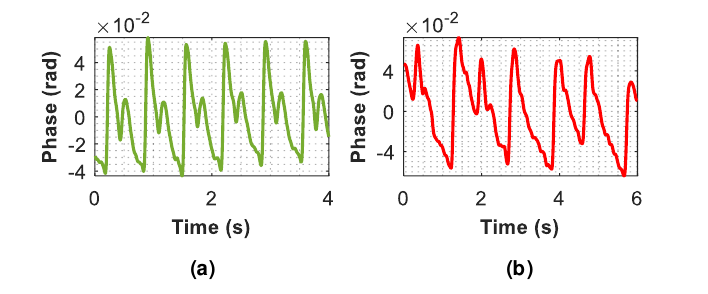}
  \caption{Comparison of CPR results between subjects with \textbf{(a)} NSR and \textbf{(b)} AF. }

  \label{f: CPR_results}
\end{figure}

Fig.~\ref{f: CPR_results} illustrates the distinct patterns of CPR in subjects with normal sinus rhythm (NSR) and those with AF. High-quality signals like Fig.~\ref{f: CPR_results}(a) and Fig.~\ref{f: CPR_results}(b) exhibit a clear and stable heartbeat pattern, where AF subjects demonstrate markedly different features than NSR subjects.   
\section{AF Detection}
\label{s: af_detection}
In this section, we detail the approach taken by \systemname for detecting AF. We begin with the explanation of the relationship between ECG and CPR collected by \systemname, followed by the peak detection and extraction of CPR features. In the end, we discuss the AF detection model that utilizes machine learning techniques.

\subsection{Relationship Between ECG and CPR}
\begin{figure}
  \centering
    \subfloat[ECG waveform]{\includegraphics[width=0.4\linewidth]{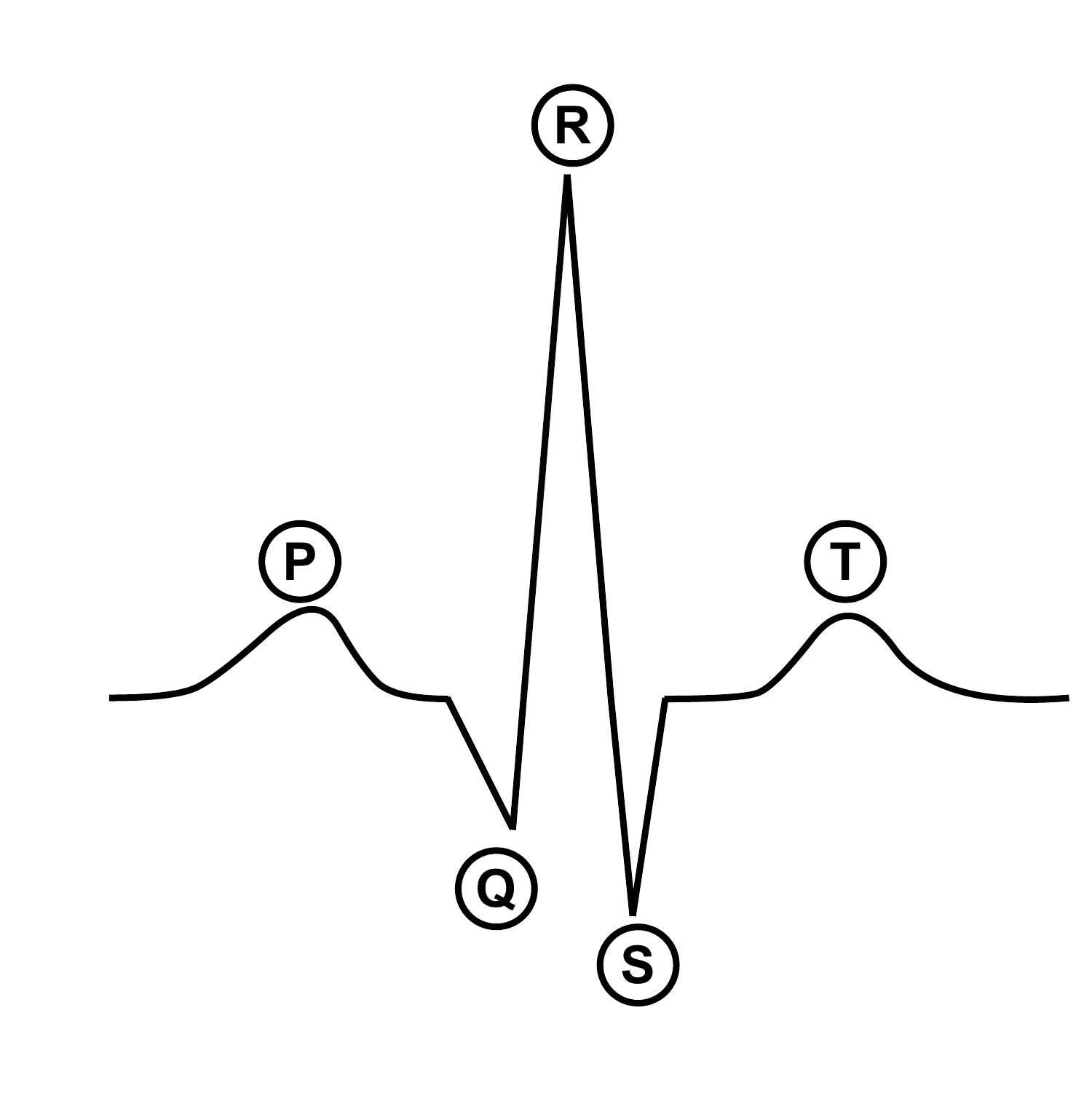}}
    \subfloat[CPR waveform]{\includegraphics[width=0.4\linewidth]{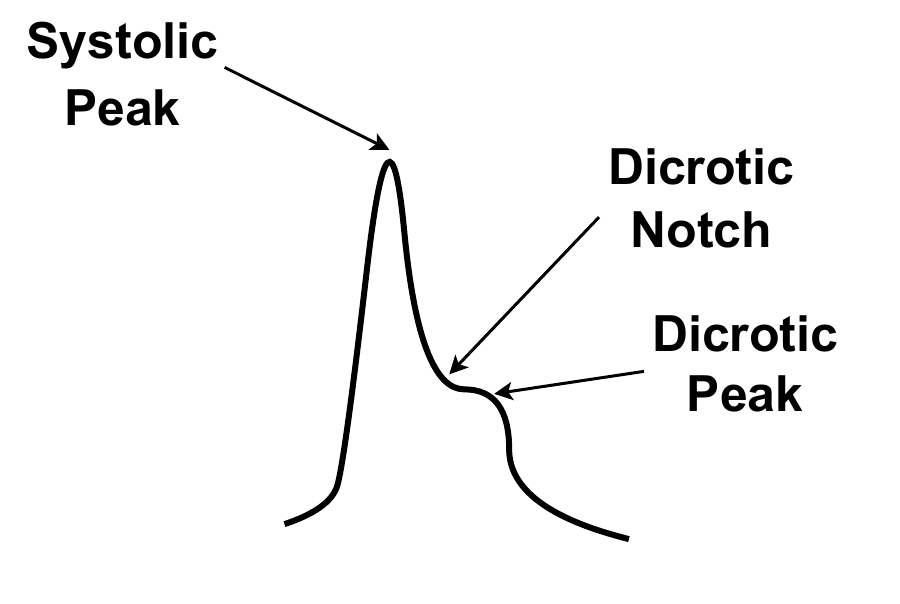}}
  \caption{ECG and CPR Waveforms}
  \label{f: ECG_CPR}
\end{figure}

\begin{figure}
\centering
    \subfloat[ECG of AF]{\includegraphics[width=0.4\linewidth]{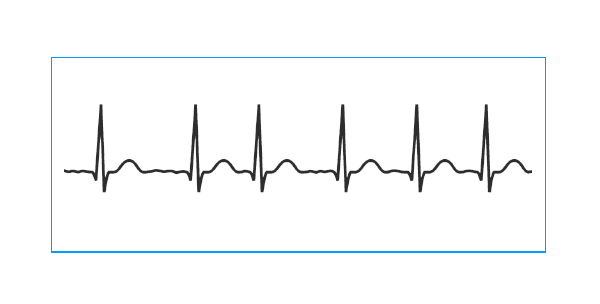}}
    \subfloat[ECG of NSR]{\includegraphics[width=0.4\linewidth]{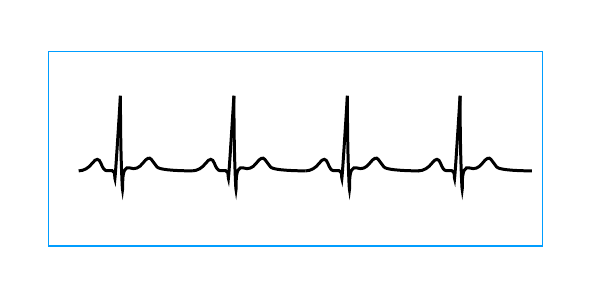}}
\caption{Comparison of ECG of AF and NSR.}
\label{f: AF_NSR}
\end{figure}

The golden method for diagnosing AF is the ECG. To implement AF detection using acoustic sensing, we have to demonstrate the relationship between the CPR waveform and the ECG waveform. ECG waveform typically consists of several components, the P wave, the T wave, and the QRS complex, as shown in Fig.~\ref{f: ECG_CPR}(a). As illustrated in Fig.~\ref{f: AF_NSR}, compared to NSR subjects, AF patients exhibit unstable intervals between adjacent R-waves in their ECG waveforms, known as RR intervals. These intervals are considered to correspond to the intervals between systolic peaks (also known as RR intervals for simplicity) in CPR waveform, as depicted in Fig.~\ref{f: ECG_CPR}(b), which contributes to differentiating AF from NSR. Hence, we extracted waveform features based on the RR intervals of CPR and used them to train the model.

\subsection{Peak Detection}
For the peak detection of systolic peaks in received signals, we used the algorithm proposed by Bishop \textit{et al.}\cite{bishop}, which is an optimization version of Scholkmann algorithm \cite{scholkmann}. The algorithm has been implemented in a python toolbox NeuronKit2 \cite{NeuroKit2} for us to utilize. 

\subsection{Feature Extraction}
We refer to the extraction method of signal features in related works \cite{feature1, feature2} and further incorporate some additional features, finally identifying the 12 RR interval features:

\begin{itemize}
    \item \textbf{minHR}: Minimum value of all RR intervals.
    \item \textbf{meanHR}: Mean value of all RR intervals.
    \item \textbf{medianHR}: Median value of all RR intervals.
    \item \textbf{skRR}: Skewness of all RR intervals.
    \item \textbf{SDRR}: Standard deviation of the RR intervals.
    \item \textbf{CVRR}: Standard deviation of the RR intervals divided by the mean of the RR intervals.
    \item \textbf{pNN50}: Proportion of RR intervals greater than 50 ms, out of the total number of RR intervals.
    \item \textbf{RMSSD}: Square root of the mean of the squared successive differences between adjacent RR intervals.
    \item \textbf{SDRMSSD}: SDRR / RMSSD, a time-domain equivalent for the low frequency to high frequency ratio \cite{feature3}.
    \item \textbf{SD ratio}: 
    \[
        \text{SD ratio} = \frac{\sqrt{0.5 \times \text{RMSSD}^2}}{\sqrt{2 \times \text{SDRR}^2 - 0.5 \times \text{RMSSD}^2}}.
    \]
    \item \textbf{SDSD}: Standard deviation of the successive differences between RR intervals.
    \item \textbf{CVSD}: Root mean square of successive differences (RMSSD) divided by the mean of the RR intervals (meanHR).
\end{itemize}

, and 14 statistical features:

\begin{itemize}
    \item \textbf{HF}: Spectral power of high frequencies (15 Hz to 4 Hz).
    \item \textbf{HFn}: Normalized high frequency, obtained by dividing the high frequency power by the total power.
    \item \textbf{LnHF}: Log-transformed HF.
    \item \textbf{TP}: Total spectral power.
    \item \textbf{SD1}: Standard deviation perpendicular to the line of identity.
    \item \textbf{SD2}: Standard deviation along the identity line. Index of long-term HRV changes.
    \item \textbf{SD1/SD2}: Ratio of SD1 to SD2.
    \item \textbf{S}: Area of ellipse described by SD1 and SD2 ($\pi \times \text{SD1} \times \text{SD2}$).
    \item \textbf{Difference}: Ratio of the sum of the differences between successive RR intervals to the sum of the differences between these differences.
    \[
        \text{Difference} = \frac{\sum_{i=0}^n \Delta y(x_i)}{\sum_{i=0}^n \Delta(\Delta y(x_i))}.
    \]
    \item \textbf{Sample Entropy}: Sample entropy of the input signal.
    \item \textbf{Shannon Entropy}: Shannon entropy of the input signal.
    \item \textbf{Approximate Entropy}: Approximate entropy of the input signal.
    \item \textbf{Multiscale Entropy}: Multiscale entropy of the input signal.
    \item \textbf{Turning Point Ratio}: Ratio of turning points (points greater or less than their two neighbors) to the total data length.
\end{itemize}

\subsection{AF Detection Model}
\label{s: af_detection_model}
Before using the extracted features as inputs for the classifier, we normalize each feature by subtracting the mean value of that feature across all samples and dividing it by the variance. This normalization process helps to mitigate the impact of the order of magnitude differences between features.

Following the normalization of features, \systemname proceeds with the machine learning-based AF detection model. We evaluate several classifiers, including Linear Support Vector Classification (SVC), AdaBoost, Decision Tree, k-Nearest Neighbors (kNN), and Random Forest. Ultimately, Linear SVC demonstrates superior performance compared to the others, which is chosen for the model training.

\section{Implementation}
\label{s: implementation}
Our system is implemented on a Redmi Note 11 Pro smartphone with built-in speakers and microphones. Acoustic data are collected using the smartphone and subsequently transferred to a computer for further processing. The smartphone records at a sampling rate of 48,000 Hz. Reference ECG data are obtained from the LEPU Three-lead ECG Monitor PC-80D, with a sample rate of 150 Hz. Both acoustic and ECG data are annotated using Label Studio, with ECG serving as the ground truth reference. Algorithms detailed in Sec.~\ref{s: pulse_wave_extraction} are implemented with MATLAB 2023b. PPG data and CPR data are aligned by their recording start time, with an error margin of less than 10 seconds. For Sec.~\ref{s: af_detection}, we use the Python library NeuroKit2\cite{NeuroKit2} to detect the peaks of pulse waves and extract features. Besides, we use Scikit-learn \cite{scikit-learn} of version 1.3.2 for model setup and training.
\section{Evaluation}
\label{s: evaluation}
In this section, we fully evaluate our system with overall performance and system robustness.

\subsection{Experiment Setup}
\subsubsection{Dataset Collection}
We recruited 20 participants aged from 20 to 89 years (average 45.5) from our university and its affiliated hospital. Among them, 6 were AF patients, and 14 were not. Experiments were conducted in a quiet conference room or ward. For evaluating \systemname's performance, we collected acoustic data using the top microphone and speakers of a Redmi Note 13 smartphone at a sample rate of 48,000 Hz, and ECG data using the LEPU three-lead ECG Monitor PC-80D at a sample rate of 150 Hz. Each data recording lasted 30 seconds, and participants engaged for about 30 minutes each. We collected 764 valid data pieces in total. The study was approved by the Institutional Review Board (IRB) of our institution.
\footnote{SUSTech IRB No. 20240084.}

\subsubsection{Ground Truth}
Since ECG is the gold standard for diagnosing atrial fibrillation, we used it as the ground truth for our data. We acquired signals at the radial artery while simultaneously using the LEPU three-lead ECG Monitor PC-80D for ECG detection. Data from the smartphones and ECG patch were aligned using timestamps. Expert cardiologists examined and labeled the records as either AF or non-AF.

\subsection{Metrics}
We used a confusion matrix to analyze our model's training results, defining AF records as positive samples and NSR records as negative samples. From the confusion matrix elements—True Negatives (TN), True Positives (TP), False Negatives (FN), and False Positives (FP)—we calculated the following metrics:

\begin{itemize}
\item \textbf{Accuracy}: Measures overall correctness, calculated as \\ $\frac{TN + TP}{TN + TP + FN + FP}$.
\item \textbf{Precision}: Quantifies correctly predicted positive samples out of all predicted positive samples, calculated as $\frac{TP}{TP + FP}$.
\item \textbf{Recall}: Assesses correctly predicted positive samples out of all actual positive samples, calculated as $\frac{TP}{TP + FN}$.
\item \textbf{F1 Score}: Considers both precision and recall, calculated as $2 \times \frac{\text{precision} \times \text{recall}}{\text{precision} + \text{recall}}$.
\end{itemize}

\begin{figure}
  \centering
  \includegraphics[width=0.6\linewidth]{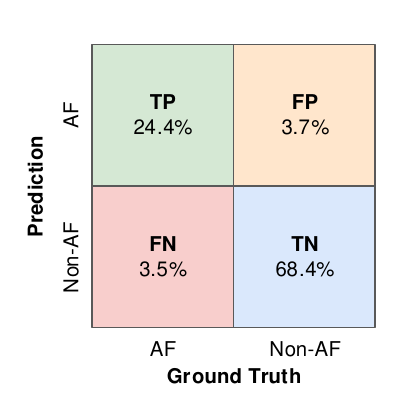}
  \caption{Averaged confusion matrix of leave-one-out validation test results.}
  \label{f: confusion_matrix}
\end{figure}
\begin{figure}
    \centering
    \includegraphics[width=0.7\linewidth]{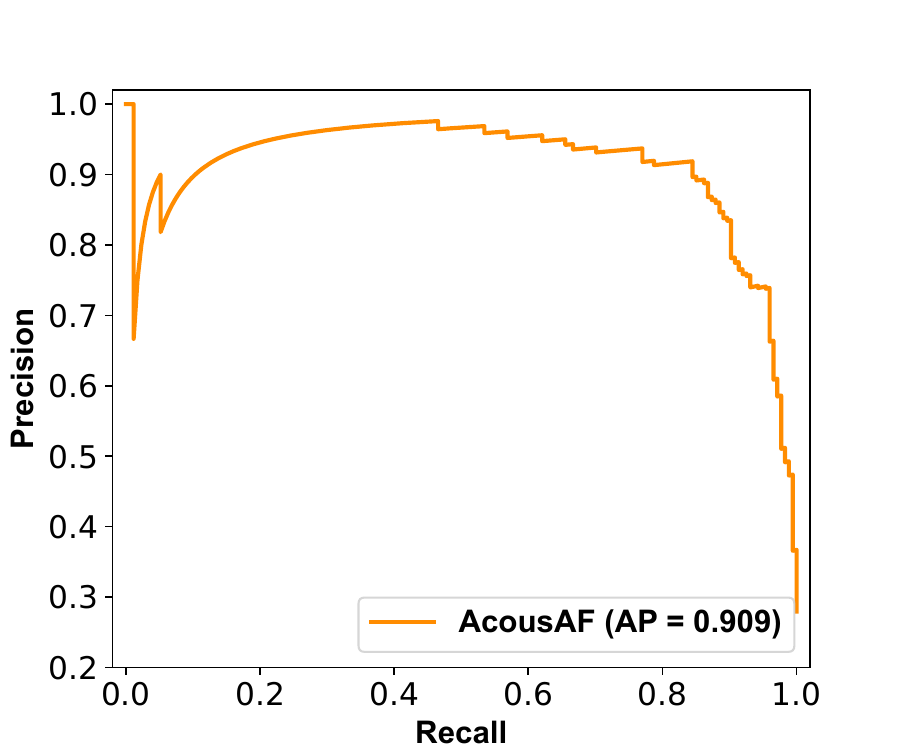}
    \caption{Precision-Recall (PR) Curve with average precision (AP).}
    \label{fig: pr-curve}
\end{figure}
\subsection{Overall Performance}
\label{s: evaluation.overall_performance}
We adopted leave-one-out cross-validation to test our model's overall performance. Each time one participant was kept as test set and the remaining participants were taken as training set. While testing each subject, the number of TP, TN, FP, and FN were recorded and accumulated for calculating performance metrics, and the decision scores were also recorded for drawing the Precision-Recall (PR) curve.

Fig.~\ref{f: confusion_matrix} illustrates the confusion matrix composed of the results from all the involved subjects. Our system achieves an average of 92.8\% accuracy, 86.9\% precision, 87.4\% recall, and 87.1\% F1 Score. That is to say, \systemname can effectively distinguish AF subjects from non-AF subjects without obvious bias. What's more, Fig.~\ref{fig: pr-curve} shows the Precision-Recall (PR) curve with average precision (AP). \systemname has an AP of 0.909, indicating that the system can accurately detect AF with relatively high recall and precision simultaneously. 

\begin{figure}
  \centering
  \includegraphics[width=0.75\linewidth]{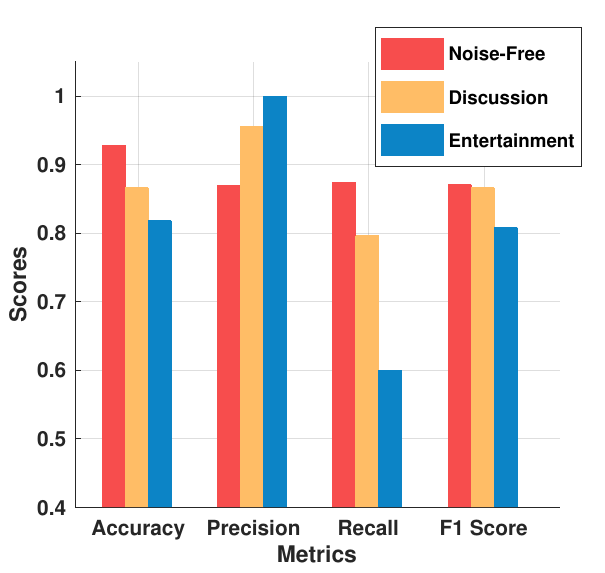}
  \caption{Performance under noise-free scenario and two noisy scenarios.}
  \label{f: noise_comparison}
\end{figure}

\subsection{Impact of Background Noise}

In our previous experiments, data were collected in noise-free environments. To evaluate the robustness of our system, we collected noise-polluted data in both conversation and entertainment scenarios as test sets. Meanwhile, the data used in the overall performance evaluation serves as the training set. We ensure that no subjects appeared in both the training and test sets. Additionally, each scenario involves one subject with AF and one without AF. In the conversation scenario, subjects were asked to talk freely at a volume exceeding 40 dB. In the entertainment scenarios, subjects were required to watch TV or play music at volumes exceeding 40 dB to simulate real-world conditions. 

Fig.~\ref{f: noise_comparison} compares the overall performance in noise-free environments to that in noisy environments, demonstrating that our system maintains relatively high performance in terms of accuracy and F1 score despite the background noise. Notably, the system shows extremely high precision and very low recall in noisy scenarios, indicating that the model behaves cautiously under these conditions. This caution may be attributed to the insufficient amount of noise-polluted data in the training set.

\begin{figure}
  \centering
  \includegraphics[width=0.75\linewidth]{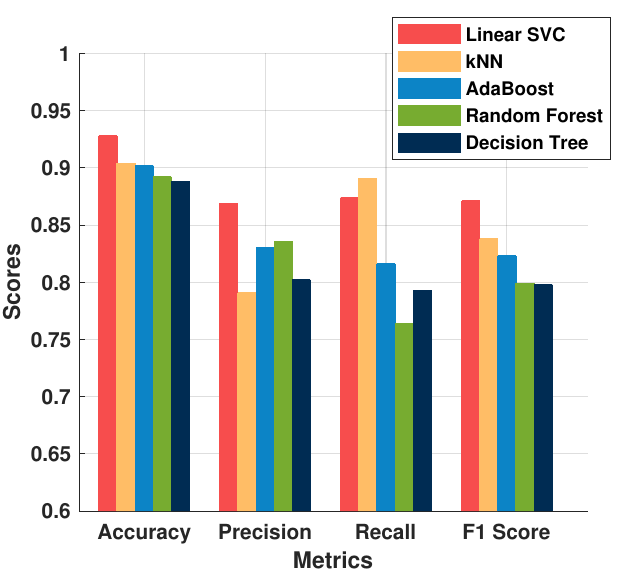}
  \caption{Performance using different classifiers.}
  \label{f: classfier_result}
\end{figure}

\subsection{Performance on Different Machine Learning Models}
As outlined in Sec.~\ref{s: af_detection_model}, \systemname constructs the AF detection model using Linear SVC. In this section, we evaluate the performance of Linear SVC in comparison with four other well-known classifiers: AdaBoost, Decision Tree, kNN, and Random Forest. The results are presented in Fig.~\ref{f: classfier_result}, which indicates that Linear SVC generally outperforms the other four classifiers. Notably, the kNN classifier shows better recall performance than Linear SVC. However, kNN's performance is worse in the remaining three metrics. Overall, the comparison underscores the robustness and effectiveness of the Linear SVC model in the context of AF detection, making it the preferred choice for \systemname.

\section{Limitations and Future Work}
\label{s
}
Despite the effectiveness of \systemname in detecting AF, it has several limitations. This section outlines these limitations and suggests areas for future improvement.

\textbf{Interference from Other Arrhythmias.} \systemname may be affected by other arrhythmias, such as atrial flutter and premature contractions, which share similar features with AF. Future work should involve studying a diverse range of arrhythmias to enhance the system's differentiation capabilities.

\textbf{Pulse Wave Purification.} In Sec.~\ref{s: evaluation.overall_performance}, \systemname achieved an accuracy of 92.8\%. However, there is potential for further performance improvement by refining the pulse wave extraction pipeline. Future efforts should focus on enhancing the recovery of pulse waves under poor signal-to-noise ratio conditions to achieve higher overall performance.

\textbf{Feature Simplification.} Currently, \systemname uses 26 features for AF detection, which is time-consuming. Future work will aim to reduce the number of features to streamline the process while maintaining accuracy.

\textbf{Privacy and Security.} \systemname leverages mobile phones' speakers and microphones to probe users' pulse waves for AF detection. However, the use of microphones may compromise users' privacy and introduce vulnerabilities. Future work will explore ways to mitigate these risks and ensure the security and privacy of users' data.
\section{Conclusion}
\label{s: conclusion}
In conclusion, this paper presents \systemname, a novel AF detection system based on acoustic sensors of COTS smartphones. In specific, \systemname firstly validates the feasibility of leveraging acoustic sensors of smartphones to acquire high-fidelity pulse waves. In addition, a well-designed processing pipeline that incorporates pulse wave probing, pulse wave extraction, and AF detection is proposed to facilitate accurate AF detection. Furthermore, we collect data from 20 participants utilizing our implemented data collection application. Extensive experiment results on these participants across diverse scenarios demonstrate that our system can achieve superior performance, with an overall performance of 92.8\% accuracy, 86.9\% precision, 87.4\% recall, and 87.1\% F1 Score in AF detection.

\begin{acks}
We gratefully acknowledge the support of the Special Funds for the Cultivation of Guangdong College Students' Scientific and Technological Innovation ("Climbing Program" Special Funds) with project No. pdjh2024c11102.
\end{acks}

\bibliographystyle{ACM-Reference-Format}
\balance
\bibliography{ref}


\begin{thebibliography}{30}


\ifx \showCODEN    \undefined \def \showCODEN     #1{\unskip}     \fi
\ifx \showDOI      \undefined \def \showDOI       #1{#1}\fi
\ifx \showISBNx    \undefined \def \showISBNx     #1{\unskip}     \fi
\ifx \showISBNxiii \undefined \def \showISBNxiii  #1{\unskip}     \fi
\ifx \showISSN     \undefined \def \showISSN      #1{\unskip}     \fi
\ifx \showLCCN     \undefined \def \showLCCN      #1{\unskip}     \fi
\ifx \shownote     \undefined \def \shownote      #1{#1}          \fi
\ifx \showarticletitle \undefined \def \showarticletitle #1{#1}   \fi
\ifx \showURL      \undefined \def \showURL       {\relax}        \fi
\providecommand\bibfield[2]{#2}
\providecommand\bibinfo[2]{#2}
\providecommand\natexlab[1]{#1}
\providecommand\showeprint[2][]{arXiv:#2}

\bibitem[Bashar et~al\mbox{.}(2019)]%
        {WristPPG-1}
\bibfield{author}{\bibinfo{person}{Syed~Khairul Bashar}, \bibinfo{person}{Dong Han}, \bibinfo{person}{Shirin Hajeb-Mohammadalipour}, \bibinfo{person}{Eric Ding}, \bibinfo{person}{Cody Whitcomb}, \bibinfo{person}{David~D. McManus}, {and} \bibinfo{person}{Ki~H. Chon}.} \bibinfo{year}{2019}\natexlab{}.
\newblock \showarticletitle{Atrial Fibrillation Detection from Wrist Photoplethysmography Signals Using Smartwatches}.
\newblock \bibinfo{journal}{\emph{Scientific Reports}} \bibinfo{volume}{9}, \bibinfo{number}{1} (\bibinfo{date}{21 Oct} \bibinfo{year}{2019}), \bibinfo{pages}{15054}.
\newblock
\showISSN{2045-2322}
\urldef\tempurl%
\url{https://doi.org/10.1038/s41598-019-49092-2}
\showDOI{\tempurl}


\bibitem[Bishop and Ercole(2018)]%
        {bishop}
\bibfield{author}{\bibinfo{person}{Steven~M Bishop} {and} \bibinfo{person}{Ari Ercole}.} \bibinfo{year}{2018}\natexlab{}.
\newblock \showarticletitle{Multi-scale peak and trough detection optimised for periodic and quasi-periodic neuroscience data}. In \bibinfo{booktitle}{\emph{Intracranial Pressure \& Neuromonitoring XVI}}. Springer, \bibinfo{pages}{189--195}.
\newblock


\bibitem[Chan and Choy(2017)]%
        {MobileECG-4}
\bibfield{author}{\bibinfo{person}{Ngai-yin Chan} {and} \bibinfo{person}{Chi-chung Choy}.} \bibinfo{year}{2017}\natexlab{}.
\newblock \showarticletitle{Screening for atrial fibrillation in 13 122 {Hong} {Kong} citizens with smartphone electrocardiogram}.
\newblock \bibinfo{journal}{\emph{Heart}} \bibinfo{volume}{103}, \bibinfo{number}{1} (\bibinfo{date}{Jan.} \bibinfo{year}{2017}), \bibinfo{pages}{24--31}.
\newblock
\showISSN{1355-6037, 1468-201X}
\urldef\tempurl%
\url{https://doi.org/10.1136/heartjnl-2016-309993}
\showDOI{\tempurl}


\bibitem[Chan et~al\mbox{.}(2018)]%
        {MobileECG-2}
\bibfield{author}{\bibinfo{person}{Ngai-Yin Chan}, \bibinfo{person}{Chi-Chung Choy}, \bibinfo{person}{Chi-Kin Chan}, {and} \bibinfo{person}{Chung-Wah Siu}.} \bibinfo{year}{2018}\natexlab{}.
\newblock \showarticletitle{Effectiveness of a nongovernmental organization–led large-scale community atrial fibrillation screening program using the smartphone electrocardiogram: {An} observational cohort study}.
\newblock \bibinfo{journal}{\emph{Heart Rhythm}} \bibinfo{volume}{15}, \bibinfo{number}{9} (\bibinfo{date}{Sept.} \bibinfo{year}{2018}), \bibinfo{pages}{1306--1311}.
\newblock
\showISSN{15475271}
\urldef\tempurl%
\url{https://doi.org/10.1016/j.hrthm.2018.06.006}
\showDOI{\tempurl}


\bibitem[Eerikäinen et~al\mbox{.}(2018)]%
        {WristPPG-5}
\bibfield{author}{\bibinfo{person}{Linda~M Eerikäinen}, \bibinfo{person}{Alberto~G Bonomi}, \bibinfo{person}{Fons Schipper}, \bibinfo{person}{Lukas R~C Dekker}, \bibinfo{person}{Rik Vullings}, \bibinfo{person}{Helma~M de Morree}, {and} \bibinfo{person}{Ronald~M Aarts}.} \bibinfo{year}{2018}\natexlab{}.
\newblock \showarticletitle{Comparison between electrocardiogram- and photoplethysmogram-derived features for atrial fibrillation detection in free-living conditions}.
\newblock \bibinfo{journal}{\emph{Physiological Measurement}} \bibinfo{volume}{39}, \bibinfo{number}{8} (\bibinfo{date}{aug} \bibinfo{year}{2018}), \bibinfo{pages}{084001}.
\newblock
\urldef\tempurl%
\url{https://doi.org/10.1088/1361-6579/aad2c0}
\showDOI{\tempurl}


\bibitem[Fan et~al\mbox{.}(2023)]%
        {acous4}
\bibfield{author}{\bibinfo{person}{Xiaoran Fan}, \bibinfo{person}{David Pearl}, \bibinfo{person}{Richard Howard}, \bibinfo{person}{Longfei Shangguan}, {and} \bibinfo{person}{Trausti Thormundsson}.} \bibinfo{year}{2023}\natexlab{}.
\newblock \showarticletitle{APG: Audioplethysmography for Cardiac Monitoring in Hearables}. In \bibinfo{booktitle}{\emph{Proceedings of the 29th Annual International Conference on Mobile Computing and Networking}}. \bibinfo{pages}{1--15}.
\newblock


\bibitem[Guo et~al\mbox{.}(2021)]%
        {feature1}
\bibfield{author}{\bibinfo{person}{Yutao Guo}, \bibinfo{person}{Hao Wang}, \bibinfo{person}{Hui Zhang}, \bibinfo{person}{Tong Liu}, \bibinfo{person}{Luping Li}, \bibinfo{person}{Lingjie Liu}, \bibinfo{person}{Maolin Chen}, \bibinfo{person}{Yundai Chen}, {and} \bibinfo{person}{Gregory~Y.H. Lip}.} \bibinfo{year}{2021}\natexlab{}.
\newblock \showarticletitle{Photoplethysmography-Based Machine Learning Approaches for Atrial Fibrillation Prediction: A Report From the Huawei Heart Study}.
\newblock \bibinfo{journal}{\emph{JACC: Asia}} \bibinfo{volume}{1}, \bibinfo{number}{3} (\bibinfo{year}{2021}), \bibinfo{pages}{399--408}.
\newblock
\showISSN{2772-3747}
\urldef\tempurl%
\url{https://doi.org/10.1016/j.jacasi.2021.09.004}
\showDOI{\tempurl}


\bibitem[Haberman et~al\mbox{.}(2015)]%
        {MobileECG-5}
\bibfield{author}{\bibinfo{person}{Zachary~C. Haberman}, \bibinfo{person}{Ryan~T. Jahn}, \bibinfo{person}{Rupan Bose}, \bibinfo{person}{Han Tun}, \bibinfo{person}{Jerold~S. Shinbane}, \bibinfo{person}{Rahul~N. Doshi}, \bibinfo{person}{Philip~M. Chang}, {and} \bibinfo{person}{Leslie~A. Saxon}.} \bibinfo{year}{2015}\natexlab{}.
\newblock \showarticletitle{Wireless {Smartphone} {ECG} {Enables} {Large}-{Scale} {Screening} in {Diverse} {Populations}}.
\newblock \bibinfo{journal}{\emph{Journal of Cardiovascular Electrophysiology}} \bibinfo{volume}{26}, \bibinfo{number}{5} (\bibinfo{year}{2015}), \bibinfo{pages}{520--526}.
\newblock
\showISSN{1540-8167}
\urldef\tempurl%
\url{https://doi.org/10.1111/jce.12634}
\showDOI{\tempurl}


\bibitem[Han et~al\mbox{.}(2020)]%
        {MobileECG-6}
\bibfield{author}{\bibinfo{person}{Dong Han}, \bibinfo{person}{Syed~Khairul Bashar}, \bibinfo{person}{Fahimeh Mohagheghian}, \bibinfo{person}{Eric Ding}, \bibinfo{person}{Cody Whitcomb}, \bibinfo{person}{David~D. McManus}, {and} \bibinfo{person}{Ki~H. Chon}.} \bibinfo{year}{2020}\natexlab{}.
\newblock \showarticletitle{Premature {Atrial} and {Ventricular} {Contraction} {Detection} {Using} {Photoplethysmographic} {Data} from a {Smartwatch}}.
\newblock \bibinfo{journal}{\emph{Sensors}} \bibinfo{volume}{20}, \bibinfo{number}{19} (\bibinfo{date}{Oct.} \bibinfo{year}{2020}), \bibinfo{pages}{5683}.
\newblock
\showISSN{1424-8220}
\urldef\tempurl%
\url{https://doi.org/10.3390/s20195683}
\showDOI{\tempurl}


\bibitem[Isakadze and Martin(2020)]%
        {ECG3}
\bibfield{author}{\bibinfo{person}{Nino Isakadze} {and} \bibinfo{person}{Seth~S Martin}.} \bibinfo{year}{2020}\natexlab{}.
\newblock \showarticletitle{How useful is the smartwatch ECG?}
\newblock \bibinfo{journal}{\emph{Trends in cardiovascular medicine}} \bibinfo{volume}{30}, \bibinfo{number}{7} (\bibinfo{year}{2020}), \bibinfo{pages}{442--448}.
\newblock


\bibitem[Koivisto et~al\mbox{.}(2015)]%
        {SCG2}
\bibfield{author}{\bibinfo{person}{Tero Koivisto}, \bibinfo{person}{Mikko P{\"a}nk{\"a}{\"a}l{\"a}}, \bibinfo{person}{Tero Hurnanen}, \bibinfo{person}{Tuija Vasankari}, \bibinfo{person}{Tuomas Kiviniemi}, \bibinfo{person}{Antti Saraste}, {and} \bibinfo{person}{Juhani Airaksinen}.} \bibinfo{year}{2015}\natexlab{}.
\newblock \showarticletitle{Automatic detection of atrial fibrillation using MEMS accelerometer}. In \bibinfo{booktitle}{\emph{2015 Computing in Cardiology Conference (CinC)}}. IEEE, \bibinfo{pages}{829--832}.
\newblock


\bibitem[Lai et~al\mbox{.}(2020)]%
        {MobileECG-1}
\bibfield{author}{\bibinfo{person}{Dakun Lai}, \bibinfo{person}{Yuxiang Bu}, \bibinfo{person}{Ye Su}, \bibinfo{person}{Xinshu Zhang}, {and} \bibinfo{person}{Chang-Sheng Ma}.} \bibinfo{year}{2020}\natexlab{}.
\newblock \showarticletitle{A {Flexible} {Multilayered} {Dry} {Electrode} and {Assembly} to {Single}-{Lead} {ECG} {Patch} to {Monitor} {Atrial} {Fibrillation} in a {Real}-{Life} {Scenario}}.
\newblock \bibinfo{journal}{\emph{IEEE Sensors Journal}} \bibinfo{volume}{20}, \bibinfo{number}{20} (\bibinfo{date}{Oct.} \bibinfo{year}{2020}), \bibinfo{pages}{12295--12306}.
\newblock
\showISSN{1530-437X, 1558-1748, 2379-9153}
\urldef\tempurl%
\url{https://doi.org/10.1109/JSEN.2020.2999101}
\showDOI{\tempurl}


\bibitem[Lau et~al\mbox{.}(2013)]%
        {ECG2}
\bibfield{author}{\bibinfo{person}{Jerrett~K Lau}, \bibinfo{person}{Nicole Lowres}, \bibinfo{person}{Lis Neubeck}, \bibinfo{person}{David~B Brieger}, \bibinfo{person}{Raymond~W Sy}, \bibinfo{person}{Connor~D Galloway}, \bibinfo{person}{David~E Albert}, {and} \bibinfo{person}{Saul~B Freedman}.} \bibinfo{year}{2013}\natexlab{}.
\newblock \showarticletitle{iPhone ECG application for community screening to detect silent atrial fibrillation: a novel technology to prevent stroke}.
\newblock \bibinfo{journal}{\emph{International journal of cardiology}} \bibinfo{volume}{165}, \bibinfo{number}{1} (\bibinfo{year}{2013}), \bibinfo{pages}{193--194}.
\newblock


\bibitem[Lemay et~al\mbox{.}(2016)]%
        {WristPPG-2}
\bibfield{author}{\bibinfo{person}{Mathieu Lemay}, \bibinfo{person}{Sibylle Fallet}, \bibinfo{person}{Philippe Renevey}, \bibinfo{person}{Josep Solà}, \bibinfo{person}{Célestin Leupi}, \bibinfo{person}{Etienne Pruvot}, {and} \bibinfo{person}{Jean-Marc Vesin}.} \bibinfo{year}{2016}\natexlab{}.
\newblock \showarticletitle{Wrist-located optical device for atrial fibrillation screening: A clinical study on twenty patients}. In \bibinfo{booktitle}{\emph{2016 Computing in Cardiology Conference (CinC)}}. \bibinfo{pages}{681--684}.
\newblock
\showISSN{2325-887X}


\bibitem[Lippi et~al\mbox{.}(2021)]%
        {AF-Intro}
\bibfield{author}{\bibinfo{person}{Giuseppe Lippi}, \bibinfo{person}{Fabian Sanchis-Gomar}, {and} \bibinfo{person}{Gianfranco Cervellin}.} \bibinfo{year}{2021}\natexlab{}.
\newblock \showarticletitle{Global epidemiology of atrial fibrillation: An increasing epidemic and public health challenge}.
\newblock \bibinfo{journal}{\emph{International Journal of Stroke}} \bibinfo{volume}{16}, \bibinfo{number}{2} (\bibinfo{year}{2021}), \bibinfo{pages}{217--221}.
\newblock
\urldef\tempurl%
\url{https://doi.org/10.1177/1747493019897870}
\showDOI{\tempurl}


\bibitem[Makowski et~al\mbox{.}(2021)]%
        {NeuroKit2}
\bibfield{author}{\bibinfo{person}{Dominique Makowski}, \bibinfo{person}{Tam Pham}, \bibinfo{person}{Zen~J. Lau}, \bibinfo{person}{Jan~C. Brammer}, \bibinfo{person}{Fran{\c{c}}ois Lespinasse}, \bibinfo{person}{Hung Pham}, \bibinfo{person}{Christopher Schölzel}, {and} \bibinfo{person}{S.~H.~Annabel Chen}.} \bibinfo{year}{2021}\natexlab{}.
\newblock \showarticletitle{{NeuroKit}2: A Python toolbox for neurophysiological signal processing}.
\newblock \bibinfo{journal}{\emph{Behavior Research Methods}} \bibinfo{volume}{53}, \bibinfo{number}{4} (\bibinfo{date}{feb} \bibinfo{year}{2021}), \bibinfo{pages}{1689--1696}.
\newblock
\urldef\tempurl%
\url{https://doi.org/10.3758/s13428-020-01516-y}
\showDOI{\tempurl}


\bibitem[Marini et~al\mbox{.}(2005)]%
        {AF-With-Stroke}
\bibfield{author}{\bibinfo{person}{Carmine Marini}, \bibinfo{person}{Federica~De Santis}, \bibinfo{person}{Simona Sacco}, \bibinfo{person}{Tommasina Russo}, \bibinfo{person}{Luigi Olivieri}, \bibinfo{person}{Rocco Totaro}, {and} \bibinfo{person}{Antonio Carolei}.} \bibinfo{year}{2005}\natexlab{}.
\newblock \showarticletitle{Contribution of Atrial Fibrillation to Incidence and Outcome of Ischemic Stroke}.
\newblock \bibinfo{journal}{\emph{Stroke}} \bibinfo{volume}{36}, \bibinfo{number}{6} (\bibinfo{year}{2005}), \bibinfo{pages}{1115--1119}.
\newblock
\urldef\tempurl%
\url{https://doi.org/10.1161/01.STR.0000166053.83476.4a}
\showDOI{\tempurl}


\bibitem[Nemati et~al\mbox{.}(2016)]%
        {WristPPG-4}
\bibfield{author}{\bibinfo{person}{Shamim Nemati}, \bibinfo{person}{Mohammad~M. Ghassemi}, \bibinfo{person}{Vaidehi Ambai}, \bibinfo{person}{Nino Isakadze}, \bibinfo{person}{Oleksiy Levantsevych}, \bibinfo{person}{Amit Shah}, {and} \bibinfo{person}{Gari~D. Clifford}.} \bibinfo{year}{2016}\natexlab{}.
\newblock \showarticletitle{Monitoring and detecting atrial fibrillation using wearable technology}. In \bibinfo{booktitle}{\emph{2016 38th Annual International Conference of the IEEE Engineering in Medicine and Biology Society (EMBC)}}. \bibinfo{pages}{3394--3397}.
\newblock
\showISSN{1558-4615}
\urldef\tempurl%
\url{https://doi.org/10.1109/EMBC.2016.7591456}
\showDOI{\tempurl}


\bibitem[Orchard et~al\mbox{.}(2016)]%
        {MobileECG-3}
\bibfield{author}{\bibinfo{person}{Jessica Orchard}, \bibinfo{person}{Nicole Lowres}, \bibinfo{person}{S~Ben Freedman}, \bibinfo{person}{Laila Ladak}, \bibinfo{person}{William Lee}, \bibinfo{person}{Nicholas Zwar}, \bibinfo{person}{David Peiris}, \bibinfo{person}{Yasith Kamaladasa}, \bibinfo{person}{Jialin Li}, {and} \bibinfo{person}{Lis Neubeck}.} \bibinfo{year}{2016}\natexlab{}.
\newblock \showarticletitle{Screening for atrial fibrillation during influenza vaccinations by primary care nurses using a smartphone electrocardiograph ({iECG}): {A} feasibility study}.
\newblock \bibinfo{journal}{\emph{European Journal of Preventive Cardiology}}  \bibinfo{volume}{23} (\bibinfo{date}{Oct.} \bibinfo{year}{2016}), \bibinfo{pages}{13--20}.
\newblock
\showISSN{2047-4873, 2047-4881}
\urldef\tempurl%
\url{https://doi.org/10.1177/2047487316670255}
\showDOI{\tempurl}


\bibitem[Pedregosa et~al\mbox{.}(2011)]%
        {scikit-learn}
\bibfield{author}{\bibinfo{person}{F. Pedregosa}, \bibinfo{person}{G. Varoquaux}, \bibinfo{person}{A. Gramfort}, \bibinfo{person}{V. Michel}, \bibinfo{person}{B. Thirion}, \bibinfo{person}{O. Grisel}, \bibinfo{person}{M. Blondel}, \bibinfo{person}{P. Prettenhofer}, \bibinfo{person}{R. Weiss}, \bibinfo{person}{V. Dubourg}, \bibinfo{person}{J. Vanderplas}, \bibinfo{person}{A. Passos}, \bibinfo{person}{D. Cournapeau}, \bibinfo{person}{M. Brucher}, \bibinfo{person}{M. Perrot}, {and} \bibinfo{person}{E. Duchesnay}.} \bibinfo{year}{2011}\natexlab{}.
\newblock \showarticletitle{Scikit-learn: Machine Learning in {P}ython}.
\newblock \bibinfo{journal}{\emph{Journal of Machine Learning Research}}  \bibinfo{volume}{12} (\bibinfo{year}{2011}), \bibinfo{pages}{2825--2830}.
\newblock


\bibitem[Puranen et~al\mbox{.}(2020)]%
        {PPGSkinTone}
\bibfield{author}{\bibinfo{person}{Antti Puranen}, \bibinfo{person}{Tuomas Halkola}, \bibinfo{person}{Ole Kirkeby}, {and} \bibinfo{person}{Antti Vehkaoja}.} \bibinfo{year}{2020}\natexlab{}.
\newblock \showarticletitle{Effect of skin tone and activity on the performance of wrist-worn optical beat-to-beat heart rate monitoring}. In \bibinfo{booktitle}{\emph{2020 IEEE SENSORS}}. \bibinfo{pages}{1--4}.
\newblock
\urldef\tempurl%
\url{https://doi.org/10.1109/SENSORS47125.2020.9278523}
\showDOI{\tempurl}


\bibitem[Pänkäälä et~al\mbox{.}(2016)]%
        {SCGAF}
\bibfield{author}{\bibinfo{person}{Mikko Pänkäälä}, \bibinfo{person}{Tero Koivisto}, \bibinfo{person}{Olli Lahdenoja}, \bibinfo{person}{Tuomas Kiviniemi}, \bibinfo{person}{Antti Saraste}, \bibinfo{person}{Tuija Vasankari}, {and} \bibinfo{person}{Juhani Airaksinen}.} \bibinfo{year}{2016}\natexlab{}.
\newblock \showarticletitle{Detection of atrial fibrillation with seismocardiography}. In \bibinfo{booktitle}{\emph{2016 38th Annual International Conference of the IEEE Engineering in Medicine and Biology Society (EMBC)}}. \bibinfo{pages}{4369--4374}.
\newblock
\urldef\tempurl%
\url{https://doi.org/10.1109/EMBC.2016.7591695}
\showDOI{\tempurl}


\bibitem[Savelieva and Camm(2000)]%
        {silentAF}
\bibfield{author}{\bibinfo{person}{Irina Savelieva} {and} \bibinfo{person}{A.~John Camm}.} \bibinfo{year}{2000}\natexlab{}.
\newblock \showarticletitle{Clinical Relevance of Silent Atrial Fibrillation: Prevalence, Prognosis, Quality of Life, and Management}.
\newblock \bibinfo{journal}{\emph{Journal of Interventional Cardiac Electrophysiology}} \bibinfo{volume}{4}, \bibinfo{number}{2} (\bibinfo{date}{01 Jun} \bibinfo{year}{2000}), \bibinfo{pages}{369--382}.
\newblock
\showISSN{1572-8595}
\urldef\tempurl%
\url{https://doi.org/10.1023/A:1009823001707}
\showDOI{\tempurl}


\bibitem[Scholkmann et~al\mbox{.}(2012)]%
        {scholkmann}
\bibfield{author}{\bibinfo{person}{Felix Scholkmann}, \bibinfo{person}{Jens Boss}, {and} \bibinfo{person}{Martin Wolf}.} \bibinfo{year}{2012}\natexlab{}.
\newblock \showarticletitle{An efficient algorithm for automatic peak detection in noisy periodic and quasi-periodic signals}.
\newblock \bibinfo{journal}{\emph{Algorithms}} \bibinfo{volume}{5}, \bibinfo{number}{4} (\bibinfo{year}{2012}), \bibinfo{pages}{588--603}.
\newblock


\bibitem[Shan et~al\mbox{.}(2016)]%
        {feature2}
\bibfield{author}{\bibinfo{person}{Shih-Ming Shan}, \bibinfo{person}{Sung-Chun Tang}, \bibinfo{person}{Pei-Wen Huang}, \bibinfo{person}{Yu-Min Lin}, \bibinfo{person}{Wei-Han Huang}, \bibinfo{person}{Dar-Ming Lai}, {and} \bibinfo{person}{An-Yeu~Andy Wu}.} \bibinfo{year}{2016}\natexlab{}.
\newblock \showarticletitle{Reliable PPG-based algorithm in atrial fibrillation detection}. In \bibinfo{booktitle}{\emph{2016 IEEE Biomedical Circuits and Systems Conference (BioCAS)}}. IEEE, \bibinfo{pages}{340--343}.
\newblock


\bibitem[Shashikumar et~al\mbox{.}(2017)]%
        {WristPPG-6}
\bibfield{author}{\bibinfo{person}{Supreeth~Prajwal Shashikumar}, \bibinfo{person}{Amit~J. Shah}, \bibinfo{person}{Qiao Li}, \bibinfo{person}{Gari~D. Clifford}, {and} \bibinfo{person}{Shamim Nemati}.} \bibinfo{year}{2017}\natexlab{}.
\newblock \showarticletitle{A deep learning approach to monitoring and detecting atrial fibrillation using wearable technology}. In \bibinfo{booktitle}{\emph{2017 IEEE EMBS International Conference on Biomedical \& Health Informatics (BHI)}}. \bibinfo{pages}{141--144}.
\newblock
\urldef\tempurl%
\url{https://doi.org/10.1109/BHI.2017.7897225}
\showDOI{\tempurl}


\bibitem[Sollers et~al\mbox{.}(2007)]%
        {feature3}
\bibfield{author}{\bibinfo{person}{John~J Sollers}, \bibinfo{person}{Tony~W Buchanan}, \bibinfo{person}{Samantha~M Mowrer}, \bibinfo{person}{LaBarron~K Hill}, {and} \bibinfo{person}{Julian~F Thayer}.} \bibinfo{year}{2007}\natexlab{}.
\newblock \showarticletitle{Comparison of the ratio of the standard deviation of the RR interval and the root mean squared successive differences (SD/rMSSD) to the low frequency-to-high frequency (LF/HF) ratio in a patient population and normal healthy controls}.
\newblock \bibinfo{journal}{\emph{Biomed Sci Instrum}}  \bibinfo{volume}{43} (\bibinfo{year}{2007}), \bibinfo{pages}{158--163}.
\newblock


\bibitem[Tarniceriu et~al\mbox{.}(2018)]%
        {WristPPG-3}
\bibfield{author}{\bibinfo{person}{Adrian Tarniceriu}, \bibinfo{person}{Jarkko Harju}, \bibinfo{person}{Zeinab~Rezaei Yousefi}, \bibinfo{person}{Antti Vehkaoja}, \bibinfo{person}{Jakub Parak}, \bibinfo{person}{Arvi Yli-Hankala}, {and} \bibinfo{person}{Ilkka Korhonen}.} \bibinfo{year}{2018}\natexlab{}.
\newblock \showarticletitle{The Accuracy of Atrial Fibrillation Detection from Wrist Photoplethysmography. A Study on Post-Operative Patients}. In \bibinfo{booktitle}{\emph{2018 40th Annual International Conference of the IEEE Engineering in Medicine and Biology Society (EMBC)}}. \bibinfo{pages}{1--4}.
\newblock
\showISSN{1558-4615}
\urldef\tempurl%
\url{https://doi.org/10.1109/EMBC.2018.8513197}
\showDOI{\tempurl}


\bibitem[Wang et~al\mbox{.}(2023)]%
        {SCG1}
\bibfield{author}{\bibinfo{person}{Lei Wang}, \bibinfo{person}{Xingwei Wang}, \bibinfo{person}{Dalin Zhang}, \bibinfo{person}{Xiaolei Ma}, \bibinfo{person}{Yong Zhang}, \bibinfo{person}{Haipeng Dai}, \bibinfo{person}{Chenren Xu}, \bibinfo{person}{Zhijun Li}, {and} \bibinfo{person}{Tao Gu}.} \bibinfo{year}{2023}\natexlab{}.
\newblock \showarticletitle{Knowing Your Heart Condition Anytime: User-Independent ECG Measurement Using Commercial Mobile Phones}.
\newblock \bibinfo{journal}{\emph{Proceedings of the ACM on Interactive, Mobile, Wearable and Ubiquitous Technologies}} \bibinfo{volume}{7}, \bibinfo{number}{3} (\bibinfo{year}{2023}), \bibinfo{pages}{1--28}.
\newblock


\bibitem[Zhang et~al\mbox{.}(2021)]%
        {EarlyDetectionAf}
\bibfield{author}{\bibinfo{person}{Hanbin Zhang}, \bibinfo{person}{Li Zhu}, \bibinfo{person}{Viswam Nathan}, \bibinfo{person}{Jilong Kuang}, \bibinfo{person}{Jacob Kim}, \bibinfo{person}{Jun~Alex Gao}, {and} \bibinfo{person}{Jeffrey Olgin}.} \bibinfo{year}{2021}\natexlab{}.
\newblock \showarticletitle{Towards Early Detection and Burden Estimation of Atrial Fibrillation in an Ambulatory Free-living Environment}.
\newblock \bibinfo{journal}{\emph{Proc. ACM Interact. Mob. Wearable Ubiquitous Technol.}} \bibinfo{volume}{5}, \bibinfo{number}{2}, Article \bibinfo{articleno}{86} (\bibinfo{date}{jun} \bibinfo{year}{2021}), \bibinfo{numpages}{19}~pages.
\newblock
\urldef\tempurl%
\url{https://doi.org/10.1145/3463503}
\showDOI{\tempurl}


\end{thebibliography}

\end{document}